\documentclass[12pt,cite,epsf,epsfig]{article}
\usepackage{epsfig}
\usepackage[dvips]{color}

\begin{document}

\title{Non-zero $\theta_{13}$ and $CP$-Violation in Inverse Neutrino Mass Matrix}
\author{Surender Verma\thanks{s\_7verma@yahoo.co.in}}
\date{\textit{Department of Physics, Government Degree College, Arki 173208, INDIA.}}
\maketitle
\begin{abstract}
The generic predictions of inverse Majorana neutrino mass matrix, $M_{\nu}^{-1}$ with texture two zeros, in the basis where charged lepton mass matrix is diagonal, have been obtained for neutrino masses and mixings, especially, $\theta_{13}$ and $CP$-violating phases. Such type of mass models are natural in the context of seesaw models with Dirac neutrino mass matrix, $M_{D}$, diagonal. Out of the fifteen possible texture two zeros patterns of the inverse neutrino mass matrix only seven are found to be compatible with the available data on neutrino masses and mixings including the latest $T\emph{2}K$ observation of non-zero $\theta_{13}$.  It is, also, found that $\theta_{13}=0$ is disallowed in the mass models investigated in the present work. While the neutrino mass matrices of type I and II are found to be necessarily $CP$ violating, type III is found to be $CP$-violating for a special case where $\theta_{23}$ lie above maximality. The atmospheric mixing angle, $\theta_{23}$ is found to dictate the possible hierarchies in type I and II neutrino mass models. For type III, additional information regarding $m_{ee}$ will be required to rule out the inverted hierarchy (IH). A maximal $\theta_{23}$ is found to be disallowed in all types of texture two zero $M_{\nu}^{-1}$ Ans$\ddot{a}$tz.
\end{abstract}

At present, in elementary particle physics, the problem of origin of quark/lepton masses and mixings is the most interesting and challenging subject of research. Especially, the neutrino sector of the Standard Model (SM) of elementary particles is one of today's most intense research field. The experimental confirmation\cite{1, 2, 3, 4, 5} of neutrino oscillations triggered assiduous efforts to measure neutrino masses and mixing angles with high accuracies never achieved before. Although the experiments have made great progress in determining the lepton mixing matrix and mass-squared differences, the origin of lepton masses and mixing is not yet understood. Several approaches have been suggested in the literature in order to understand
the phenomenology of the Yukawa sector which include radiative mechanisms\cite{6}, horizontal discrete\cite{7} and continuous global/local gauge symmetries\cite{8}. All the information regarding lepton masses and mixing is encoded in the lepton mass matrices. In the basis  where charged lepton mass matrix, $M_{l}$ is diagonal, the complex neutrino mass matrix, assuming neutrinos to be the Majorana particle, contain nine free parameters\cite{9}: three neutrino mass eigenvalues ($m_1, m_2, m_3$), three mixing angles describing the mixing between mass and flavor eigenstates of neutrinos ($\theta_{12}, \theta_{23}, \theta_{13}$) and three $CP$-violating phases ($\delta, \alpha, \beta$) where $\delta$ is Dirac-type $CP$ violating phase and the additional two phases $\alpha$, $\beta$ are associated with the Majorana character of the neutrinos. The observation of the neutrinoless double beta decay\cite{10, 11, 12} and
the measurement of the corresponding decay rate with a sufficient accuracy\cite{13}, would not only be a proof that the total lepton charge is not conserved in nature, but might
provide, also, information on the type of the neutrino mass spectrum, the absolute scale of neutrino masses and the values of the Majorana $CP$-violation phases. The experimental observations reveal that not all of the parameters of the neutrino mass matrix have been measured making it impossible to fully reconstruct the neutrino mass matrix in terms of the known parameters only. To describe the observed masses and mixing angles and to elucidate the possible origin of fermion mass generation one has to reduce the number of free parameters in the Yukawa sector. This can be achieved either by applying certain flavor symmetries or employ phenomenological approaches such as texture zeros\cite{9, 14, 15, 16, 17, 18, 19}, hybrid textures\cite{20,21}, requirement of zero determinant\cite{22}, vanishing minors\cite{23} and scaling\cite{24,25} etc. in the neutrino mass matrix and obtain the predictions for the unknown parameters. Such phenomenological approaches are bound to play important role in understanding the underlying dynamics of the fermion mass generation. In particular, it is possible to enforce texture zeros, studied extensively in the literature\cite{9, 14, 15, 16, 17, 18, 19}, in arbitrary entries of the fermion mass matrices by means of a discrete Abelian flavor symmetry\cite{26}. Recently, texture two zeros of the neutrino mass matrix, in the flavor basis, have been realized within the framework of type (I+II) seesaw mechanism natural to $SO(10)$ grand unification\cite{27}. In the context of seesaw mechanism\cite{28, 29}, the neutrino mass matrix, $M_{\nu}\approx -M_{D}M_{R}^{-1}M_{D}^{T}$, where $M_{D}$ and $M_{R}$ are Dirac and right-handed Majorana neutrino mass matrices, respectively. In the basis where $M_{D}$ is diagonal the texture zeros in different entries in $M_{R}$ are identical to the texture zeros in $M_{\nu}^{-1}$\cite{30}. Such type of models, where both the charged lepton and Dirac neutrino mass matrix is diagonal, are quite natural in the context of Grand Unified Theories (GUTs). One can embed fermion mass textures with texture zeros into renormalizable field theories making the texture zero Ans$\ddot{a}$tze more credible\cite{26}. The texture two zeros in $M_{\nu}^{-1}$ can be obtained in the context of seesaw mechanism, by introducing an Abelian symmetry with one or more scalar singlets\cite{31}. There are fifteen possible texture two zero patterns for $M_{\nu}^{-1}$ out of which only seven are consistent with the neutrino data. Four out of seven viable patterns of texture two zeros in $M_{\nu}^{-1}$ are found to be equivalent to the  four patterns of texture two zero  in $M_{\nu}$ and have identical phenomenological consequences for neutrino masses and mixings\cite{31}. In the present work, we focus on the phenomenological consequences of the remaining three viable texture two zeros in $M_{\nu}^{-1}$ namely type I, II and III. In light of the recent observation from Tokai-to-Kamioka ($T\emph{2}K$) experiment of non-zero reactor angle, $\theta_{13}$, at 2.5 $\sigma$ C.L.\cite{32} and knowing the fact that most of the mass models with tri-bimaximal mixing at the leading order cannot predict such a large mixing, such type of phenomenological analyses are very important as they have high predictability and testability. The observed non-zero $\theta_{13}$ at 2.5 $\sigma$ C.L. gives an indication and possible measurement of $CP$-violation in the leptonic sector and have important implications for neutrino physics\cite{33, 34}, in general. In the present work, we have, also, studied the $CP$-violation and obtained interesting implications for Jarlskog rephasing invariant quantity $J_{CP}$, a measure of $CP$-violation in the leptonic sector and Majorana phases $\alpha$, $\beta$.

The three viable patterns of texture two zeros in $M_{\nu}^{-1}$, which have distinguishing phenomenological implications from texture two zeros in $M_{\nu}$, are:

\begin{equation}
\left(M_{\nu}^{-1}\right)^{I}=\left(
\begin{array}{ccc}
A & 0 & C \\ 0 & D & E \\ C & E & 0
\end{array}
\right),
 \left(M_{\nu}^{-1}\right)^{II}=\left(
\begin{array}{ccc}
A & B & 0 \\ B & 0 & E \\ 0 & E & F
\end{array}
\right),
\left(M_{\nu}^{-1}\right)^{III}=\left(
\begin{array}{ccc}
A & B & C \\ B & 0 & E \\ C & E & 0
\end{array}
\right)
 \end{equation}

 In the charged lepton basis, the complex symmetric mass matrix, $M_{\nu}$
can be diagonalized by a complex unitary matrix $V$:
\begin{equation}
M_{\nu}=VM_{\nu}^{diag}V^{T}
\end{equation}
where $M_{\nu}^{diag}=Diag \{m_1,m_2,m_3\}$ is the diagonal
neutrino mass matrix. The neutrino mixing matrix $V$\cite{35} can
be written as
\begin{equation}
V\equiv U P=\left(
\begin{array}{ccc}
c_{12}c_{13} & s_{12}c_{13} & s_{13}e^{-i\delta} \\
-s_{12}c_{23}-c_{12}s_{23}s_{13}e^{i\delta} &
c_{12}c_{23}-s_{12}s_{23}s_{13}e^{i\delta} & s_{23}c_{13} \\
s_{12}s_{23}-c_{12}c_{23}s_{13}e^{i\delta} &
-c_{12}s_{23}-s_{12}c_{23}s_{13}e^{i\delta} & c_{23}c_{13}
\end{array}
\right)\left(
\begin{array}{ccc}
1 & 0 & 0 \\ 0 & e^{i\alpha} & 0 \\ 0 & 0 & e^{i(\beta+\delta)}
\end{array}
\right),
\end{equation}
where $s_{ij}=\sin\theta_{ij}$ and $c_{ij}=\cos\theta_{ij}$. The
matrix $V$ is called the neutrino mixing matrix or PMNS matrix.
The matrix $U$ is the lepton analogue of the CKM quark mixing
matrix and the phase matrix $P$ contains the two Majorana phases.
Therefore the neutrino mass matrix can be written as
\begin{equation}
M_{\nu}=U P M_{\nu}^{diag}P^{T}U^{T}.
\end{equation}
The elements of the inverse neutrino mass matrix $M_{\nu}^{-1}$ can be obtained after inverting Eqn. (4). 

The texture two zeros at different entries in $M_{\nu}^{-1}$ give four real constraining equation involving the nine free parameters of the neutrino mass matrix. We solve these four equations for mass ratios ($\frac{m_{1}}{m_{2}}$, $\frac{m_{1}}{m_{3}}$) and Majorana phases ($\alpha, \beta$) in terms of the remaining free parameters ($\theta_{12}, \theta_{23}, \theta_{13}, \delta$) of the model for type I, II and III texture two zeros inverse neutrino mass matrices:

\begin{equation}
 \left.\begin{array}{c}
\frac{m_1}{m_2}e^{-2i\alpha}=-\frac{c_{12} c_{23} s_{12} e^{i \delta } \left(c_{13}^2 c_{23}^2-2 s_{13}^2
   s_{23}^2\right)+c_{12}^2 c_{23}^2 s_{13} s_{23}+s_{12}^2 s_{13} s_{23}^3 e^{2 i \delta
   }}{c_{12}^2 s_{13} s_{23}^3 e^{2 i \delta }-c_{12} c_{23} s_{12} e^{i \delta } \left(c_{13}^2
   c_{23}^2-2 s_{13}^2 s_{23}^2\right)+c_{23}^2 s_{12}^2 s_{13} s_{23}} \\

\frac{m_1}{m_3}e^{-i(2\beta+\delta)}=\frac{s_{13} s_{23}^3 e^{i \delta } \left(c_{12}^2-s_{12}^2\right)+c_{12} c_{23} s_{12} s_{23}^2
   e^{2 i \delta }+c_{12} c_{23} s_{12} s_{13}^2 \left(s_{23}^2+1\right)}{c_{12}^2 s_{13} s_{23}^3
   e^{2 i \delta }-c_{12} c_{23} s_{12} e^{i \delta } \left(c_{13}^2 c_{23}^2-2 s_{13}^2
   s_{23}^2\right)+c_{23}^2 s_{12}^2 s_{13} s_{23}}   
\end{array}  \right\} I,
\end{equation}

\begin{equation}
 \left.\begin{array}{c}
\frac{m_1}{m_2}e^{-2i\alpha}=-\frac{c_{13}^2 s_{12} e^{i \delta } \left(-c_{12} s_{23}^3+c_{23}^3 s_{12} s_{13} e^{i \delta
   }\right)+c_{23} s_{13} \left(c_{12} s_{23}+c_{23} s_{12} s_{13} e^{i \delta
   }\right){}^2}{c_{12}^2 c_{23}^3 s_{13} e^{2 i \delta }+c_{12} s_{12} s_{23} e^{i \delta }
   \left(c_{13}^2 s_{23}^2-2 c_{23}^2 s_{13}^2\right)+c_{23} s_{12}^2 s_{13} s_{23}^2}\\

\frac{m_1}{m_3}e^{-i(2\beta+\delta)}=\frac{c_{23}^3 s_{13} e^{i \delta } \left(c_{12}^2-s_{12}^2\right)-c_{12} c_{23}^2 s_{12} s_{23}
   e^{2 i \delta }-c_{12} \left(c_{23}^2+1\right) s_{12} s_{13}^2 s_{23}}{c_{12}^2 c_{23}^3 s_{13}
   e^{2 i \delta }+c_{12} s_{12} s_{23} e^{i \delta } \left(c_{13}^2 s_{23}^2-2 c_{23}^2
   s_{13}^2\right)+c_{23} s_{12}^2 s_{13} s_{23}^2}  
\end{array}  \right\} II,
\end{equation}

\begin{equation}
 \left.\begin{array}{c}
\frac{m_1}{m_2}e^{-2i\alpha}=\frac{s_{12} \left(c_{12} s_{13}+s_{12} e^{i \delta }\right)}{c_{12} \left(s_{12} s_{13}-c_{12} e^{i
   \delta }\right)} \\

\frac{m_1}{m_3}e^{-i(2\beta+\delta)}=\frac{s_{13} \left(c_{12} s_{12} s_{13}^2 e^{-i \delta }+c_{12} s_{12} e^{i \delta }-s_{13}
   \left(c_{12}^2-s_{12}^2\right)\right)}{c_{12} c_{13}^2 \left(s_{12} s_{13}-c_{12} e^{i \delta
   }\right)}  
\end{array}  \right\} III.
\end{equation}

The experimental data on neutrino masses and mixings, including the latest $T\emph{2}K$ and $MINOS$ observation of non-zero $\theta_{13}$, used in the present analysis is given in Table 1. The two values of $m_1$ calculated from Eqns. (5), (6) and (7) for each type of inverse neutrino mass matrices, respectively, using the two mass-squared differences, must be equal within the accuracy to which the present neutrino oscillation parameters are determined. 

In Table 2. we have given the Taylor series expansion of mass ratios $\left(\frac{m_1}{m_2}, \frac{m_1}{m_3}\right)$ and Majorana phase $(\alpha, \beta)$ for type I, II and Type III inverse neutrino mass matrices with texture two zeros. However, we have not used these mass ratios and Majorana phases in our analysis which is completely based on the exact Eqns. (5), (6) and (7). They are tabulated here just for the sake of illustration and to comprehend the interesting features of the correlation plots. A non-zero $\theta_{13}$ is a generic prediction of this class of models which is, also, implicit from Eqns. (5), (6) and (7). Because for $\theta_{13}=0$, $m_1$ becomes equal to $m_2$ in type I and II mass models which contradicts solar mass hierarchy. In type III mass model, $\theta_{13}=0$ is disallowed because it predicts that $m_1=0$.

In numerical analysis, we have randomly generated the input parameters of the model with the central and error values given in Table 1 to give the predictions for the unknown parameters such as the neutrino mass ratios $\left(\frac{m_1}{m_2},\frac{m_1}{m_3}\right)$ and Majorana phases ($\alpha$, $\beta$). We have, also, calculated the effective Majorana neutrino mass, $m_{ee}$ given by

\begin{equation}
m_{ee}=\left|m_1 c_{12}^2 c_{13}^2+m_2 s_{12}^2 c_{13}^2 e^{2i\alpha}+m_3 s_{13}^2 e^{2i\beta}\right|
\end{equation}
 for type I, II and III inverse neutrino mass matrices with texture two zeros. A non-zero $\theta_{13}$ in these models with texture two zero gives a hint of possible $CP$ violation in the leptonic sector as $\theta_{13}$ and the Dirac type $CP$-violating phase, $\delta$ appears together in the lepton mixing matrix. We calculate the Jarlskog rephasing invariant quantity $J_{CP}$ which is given by 
 \begin{equation}
 J_{CP}=s_{12} c_{12}s_{23}c_{23}s_{13}c_{13}^2\sin\delta
 \end{equation} 
 and is the measure of $CP$-violation in the leptonic sector.

The mass ratios $\left(\frac{m_1}{m_2}, \frac{m_1}{m_3}\right)$, upto first order in $s_{13}$, can be written as
\begin{equation}
 \left.\begin{array}{c}
\frac{m_1}{m_2}\approx1+\frac{s_{23}}{c_{12}s_{12}c_{23}^{3}}s_{13}\cos{\delta} \\

\frac{m_1}{m_3}\approx\frac{s_{23}^{2}}{c_{23}^{2}}+\frac{c_{12}s_{23}^{3}}{s_{12}c_{23}^{5}} s_{13}\cos{\delta}  
\end{array}  \right\} I,
\end{equation}

\begin{equation}
 \left.\begin{array}{c}
\frac{m_1}{m_2}\approx1-\frac{c_{23}}{c_{12}s_{12}s_{23}^{3}}s_{13}\cos{\delta}\\

\frac{m_1}{m_3}\approx\frac{c_{23}^{2}}{s_{23}^{2}}-\frac{c_{12}c_{23}^{3}}{s_{12}s_{23}^{5}} s_{13}\cos{\delta}  
\end{array}  \right\} II,
\end{equation}

\begin{equation}
 \left.\begin{array}{c}
\frac{m_1}{m_2}\approx\frac{s_{12}^{2}}{c_{12}^{2}}- \frac{s_{12}\tan{2 \theta_{23}}}{c_{12}^{3}} s_{13} \cos{\delta} \\

\frac{m_1}{m_3}\approx\tan{2 \theta_{23}} \tan{\theta_{12}} s_{13} 
\end{array}  \right\} III.
\end{equation}

From these approximate expressions of mass ratios it is implicit that $\theta_{13}=0$ is disallowed, as discussed earlier. Moreover, $\cos{\delta}$ should be negative (positive) i.e. $90^o<\delta<270^o$ ($0^o<\delta<90^o, 270^o<\delta<360^o$), as implied by the mass ratio $\frac{m_1}{m_2}$ for type-I (type-II) inverse neutrino mass matrix, in order to comply with the solar mass hierarchy. Furthermore, from Eqn. (12), it is clear that the second term of the mass ratio $\frac{m_1}{m_2}$ must always be positive otherwise $\frac{m_1}{m_2}$ may become greater than $1$ contradicting the solar mass hierarchy. So, if $\theta_{23}$ is below(above) maximality then $\cos{\delta}$ should be positive(nagative) i.e. $0^o<\delta<90^o$ or $270^o<\delta<360^o$ ($90^o<\delta<270^o$) for type-III inverse neutrino mass matrices. In the zeroth order approximation, Eqn. (10) reveals that $\theta_{23}$ should be below(above) maximality for type-I inverse neutrino mass matrices with normal(inverted) hierarchy. Similarly, $\theta_{23}$ should be above(below) maximality for normal(inverted) hierarchy for type-II inverse neutrino mass matrices. However, both regions above as well as below maximality are allowed for either hierarchies(NH and IH) of neutrino masses for type-III inverse neutrino mass matrices. We have, also, calculated the parameter $R_{\nu}\equiv\frac{\Delta m_{12}^2}{\Delta m_{23}^2}$ for type-I, II and III inverse neutrino mass matrices and can be written as

\begin{equation}
R_{\nu}^I\approx\frac{2 s_{23}^{5}}{c_{12}s_{12}c_{23}^{3}(s_{23}^2-c_{23}^2)}s_{13}\cos{\delta},
\end{equation}
\begin{equation}
R_{\nu}^{II}\approx\frac{2 c_{23}^{5}}{c_{12}s_{12}s_{23}^{3}(c_{23}^2-s_{23}^2)}s_{13}\cos{\delta},
\end{equation}

\begin{equation}
R_{\nu}^{III}\approx\frac{\tan^2{2 \theta_{23}}}{\sin{2\theta_{12}}\tan{2\theta_{12}}}s_{13}^{2}.
\end{equation}
 From Eqn. (13-15) it is evident that a maximal $\theta_{23}$ is disallowed as it imply $R_{\nu}$ to be infinite. In the following, we have presented the numerical results based on the exact Eqns.(5-7) which can be easily comprehended with the help of approximate analytical expressions (Eqns. (10-15)) and the subsequent discussion.    

 In Figure 1, we have depicted the scatter plots of $\theta_{13}$ and $\theta_{23}$ with $J_{CP}$ and Dirac type $CP$-violating phase, $\delta$, respectively for NH of neutrino masses. It is implicit from Figure 1 that $J_{CP}=0$ is disallowed for type I and II neutrino mass models and are thus, necessarily $CP$ violating. For neutrino mass model of type III, $\theta_{13}=0$ is disallowed, as discussed earlier, but $\delta=0^o$ or $360^o$ is still allowed which impel $J_{CP}$ (Eqn. (9)) to vanish in the region where $\theta_{23}$ is below maximality. However, this case becomes necessarily $CP$-violating for the region where $\theta_{23}$ is above maximality. The Dirac type $CP$-violating phase, $\delta$ is constrained to lie between $90^o$ to $270^o$ for type I with $\delta=180^o$ disallowed by the current data. For type II, $\delta$ is highly constrained to lie around $90^o$ or $270^o$. The range $90^o<\delta<270^o$ is disallowed for $\theta_{23}$ below maximality, however, for $\theta_{23}$ above maximal $\delta$ becomes highly constrained to lie in the range $90^o<\delta<270^o$, though, $\delta=180^o$ is disallowed.

The scatter plots in Figure 2 illustrate the correlation of $\theta_{23}$ with $\frac{m_1}{m_2}$ for normal as well as inverted mass hierarchy. A maximal $\theta_{23}$ is disallowed in texture two zeros inverse neutrino mass matrices (Eqns. (13-15)) as it will lead to $m_1=m_2$ which contradicts the solar mass hierarchy. Moreover, for neutrino mass models of type I (type II), $\theta_{23}$ lies below (above) maximality for NH (NH) and above (below) maximality for IH (IH). In other words, the knowledge of, whether $\theta_{23}$ is above or below maximality, will determine the type of neutrino mass hierarchy for type I and II mass matrices. $\theta_{23}$ can attain values above or below maximality in type III neutrino mass models, however, a maximal $\theta_{23}$ is still disallowed.

Majorana type $CP$-violating phases are found to be highly constrained in this class of mass models except type III where full range of $\beta$, $-90^o<\beta<90^o$, is allowed. In Figure 3, we have demonstrated the ($\alpha$-$\beta$) correlation plots for normal as well as inverted hierarchy. The effective Majorana mass, $m_{ee}$ (Eqn. (8)), depends on a number of known and unknown neutrino parameters, and testing or cross-checking the values of these parameters is obviously of immense importance\cite{36, 37}. Among the unknown neutrino parameters, the possible neutrino mass hierarchy is of particular interest. There exist a unique possibility of ruling out the inverted hierarchy (IH) of neutrino with neutrinoless double beta decay experiments because, in this case, the effective Majorana mass, $m_{ee}$ is bounded from below\cite{13}. There are a number of new experiments which may probe $m_{ee}$ at the level of $10$ meV to $50$ meV\cite{38, 39, 40, 41} and will test the phenomenological predictions of these mass models. Apart from the observance of non-zero $\theta_{13}$ in the $T\emph{2}K$ experiment, we expect occurrence of new, very interesting results soon,  since some of these $0\nu\beta\beta$ projects will start data collection in $2011-12$. In Figure 4, we have depicted the scatter plot of effective Majorana mass, $m_{ee}$ with $\theta_{23}$ for NH as well as IH of neutrino masses. It is implicit from Figure 4 that there exist a lower bound on effective Majorana neutrino mass $m_{ee}$ of about $0.05$ eV. A non-observance of $0\nu\beta\beta$ decay down to the sensitivity level of $0.05$ eV, which is achievable in the future $0\nu\beta\beta$ experiments, will rule out inverted hierarchy (IH) for type I, II and III texture two zeros inverse neutrino mass models.

In conclusions, we have presented a detailed analysis of inverse neutrino mass matrices with two zeros. The neutrino mass matrix as such contain large number of free parameters making it impossible to fully reconstruct the mass matrix, however, it is found that these mass models results in the reduction of free parameters by the imposition of texture zero which can be realized, in the context of seesaw mechanism, by introducing an Abelian symmetry with one or more scalar singlets\cite{31}. Thus, elevating the predictive power of the model. A non-zero $\theta_{13}$ is a generic prediction of these mass models. The atmospheric mixing angle, $\theta_{23}$ is found to dictate the possible hierarchies in type I and II neutrino mass models. For type III, additional information regarding $m_{ee}$ will be required to rule out the inverted hierarchy. The type I and II inverse neutrino mass matrices are found to be necessarily $CP$ violating, however, type III become $CP$-violating for a special case where $\theta_{23}$ lie above maximality. A maximal $\theta_{23}$ is found to be disallowed in all types of texture two zero $M_{\nu}^{-1}$ Ans$\ddot{a}$tz considered in the present study. It is of immense importance to note that lower bound on $m_{ee}>0.05 $ eV for IH is achievable in the future $0\nu\beta\beta$ experiments and the non-observance of which will rule out inverted hierarchy (IH) for type I, II and III mass models.

\newpage

\begin{table} [t]
\begin{center}
\begin{tabular}{cccc}
  \hline\hline
    Parameters & Best fit $\pm 1\sigma$ & $2\sigma$ & $3\sigma$ \\ \hline\hline
  $\Delta m^2_{21} [10^{-5} eV^{2}]$ & $7.58_{-0.26}^{+0.22}$ & $7.16-7.99$ & $6.99-8.18$ \\
  $|\Delta m^2_{31}| [10^{-3} eV^{2}]$ & $2.35_{-0.09}^{+0.12}$ & $2.17-2.57$ & $2.06-2.67$ \\
  $\sin^2 \theta_{12}$ & $0.312_{-0.016}^{+0.017}$ & $0.280-0.347$ & $0.265-0.364$ \\
  $\sin^2 \theta_{23}$ & $0.42_{-0.03}^{+0.08}$ & $0.36-0.60$ & $0.34-0.64$ \\
  $\sin^2 \theta_{13}$ & $0.025_{-0.007}^{+0.007}$ & $0.012-0.041$ & $0.005-0.050$ \\
  \hline
\end{tabular}
\caption{Best-fit values with $1 \sigma$ errors, $2\sigma$ and $3\sigma$ intervals for the three flavors neutrino oscillation parameters from global neutrino data analysis\cite{42}}.
\end{center}
\end{table}

\begin{table}[t]
\begin{center}
\begin{tabular}{||c|c|c||}
 \hline
     &Mass ratios & Majorana phases   \\
 \hline\hline 
 
 I&$\begin{array}{c}
\frac{m_1}{m_2}\approx\left|1+\frac{s_{13} s_{23} e^{-i \delta } \left(c_{23}^2+s_{23}^2 e^{2 i \delta
   }\right)}{c_{12} c_{23}^3 s_{12}}\right|   \\ \frac{m_1}{m_3}\approx\left|\frac{s_{23}^2}{c_{23}^2}+\frac{c_{12} s_{13} s_{23}^3 e^{-i \delta } \left(c_{23}^2+s_{23}^2 e^{2
   i \delta }\right)}{c_{23}^5 s_{12}}\right| 
\end{array}$ & $\begin{array}{c}
\alpha\approx-\frac{1}{2}Arg\left(1  + \frac{s_{13} s_{23} e^{-i \delta} \left(c_{23}^2+s_{23}^2 e^{2 i \delta }\right)}{c_{12} c_{23}^3
   s_{12}}\right)   \\ \beta\approx-\frac{1}{2}\left(Arg\left(-\frac{c_{12} s_{13} s_{23}^3 \left(c_{23}^2+s_{23}^2 e^{2 i \delta }\right)}{c_{23}^5 s_{12}}  -\frac{s_{23}^2 e^{i \delta }}{c_{23}^2}\right)+\delta \right) 
\end{array}$ \\ \hline 

 II&$\begin{array}{c}
\frac{m_1}{m_2}\approx\left|1-\frac{c_{23} s_{13} e^{-i \delta } \left(s_{23}^2+c_{23}^2 e^{2 i \delta }\right)}{c_{12} s_{12}
   s_{23}^3}\right|   \\ \frac{m_1}{m_3}\approx\left|\frac{c_{23}^2}{s_{23}^2}-\frac{c_{12} c_{23}^3 s_{13} e^{-i \delta } \left(s_{23}^2+c_{23}^2 e^{2 i
   \delta }\right)}{s_{12} s_{23}^5}\right| 
\end{array}$ & $\begin{array}{c}
\alpha\approx-\frac{1}{2}Arg\left(1-\frac{c_{23} s_{13} e^{-i \delta } \left(s_{23}^2+c_{23}^2 e^{2 i \delta }\right)}{c_{12} s_{12}
   s_{23}^3}\right)   \\ \beta\approx-\frac{1}{2}\left(Arg\left(\frac{c_{12} c_{23}^3 s_{13} \left(s_{23}^2+c_{23}^2 e^{2 i \delta }\right)}{s_{12}
   s_{23}^5}-\frac{c_{23}^2 e^{i \delta }}{s_{23}^2}\right)+\delta \right) 
\end{array}$ \\ \hline

III&$\begin{array}{c}
\frac{m_1}{m_2}\approx\left|\frac{s_{12}^2}{c_{12}^2}-\frac{2 c_{23} s_{13} s_{23} s_{12} e^{-i \delta }}{c_{12}^3
   \left(c_{23}^2-s_{23}^2\right)}\right|   \\ 
   \frac{m_1}{m_3}\approx\frac{2 c_{23} s_{12} s_{13} s_{23}}{c_{12} \left(c_{23}^2-s_{23}^2\right)} 
\end{array}$ & $\begin{array}{c}
\alpha\approx-\frac{1}{2}Arg\left(\frac{s_{12}^2}{c_{12}^2}-\frac{2 c_{23} s_{13} s_{23} s_{12} e^{-i \delta }}{c_{12}^3
   \left(c_{23}^2-s_{23}^2\right)}\right)   \\ \beta\approx-\frac{1}{2}Arg\left(\frac{2 c_{23} s_{12} s_{13} s_{23} e^{i \delta }}{c_{12} \left(c_{23}^2-s_{23}^2\right)}\right) 
\end{array}$ \\ \hline \hline

\end{tabular}
\caption{The mass ratios $\left(\frac{m_1}{m_2}, \frac{m_1}{m_3}\right)$ and Majorana phases $(\alpha, \beta)$ upto first order in the smallest leptonic mixing angle, $\theta_{13}$, for type I, II and III texture two zeros in $M_{\nu}^{-1}$.}
\end{center}
\end{table}

\begin{figure}
\begin{center}
{\epsfig{file=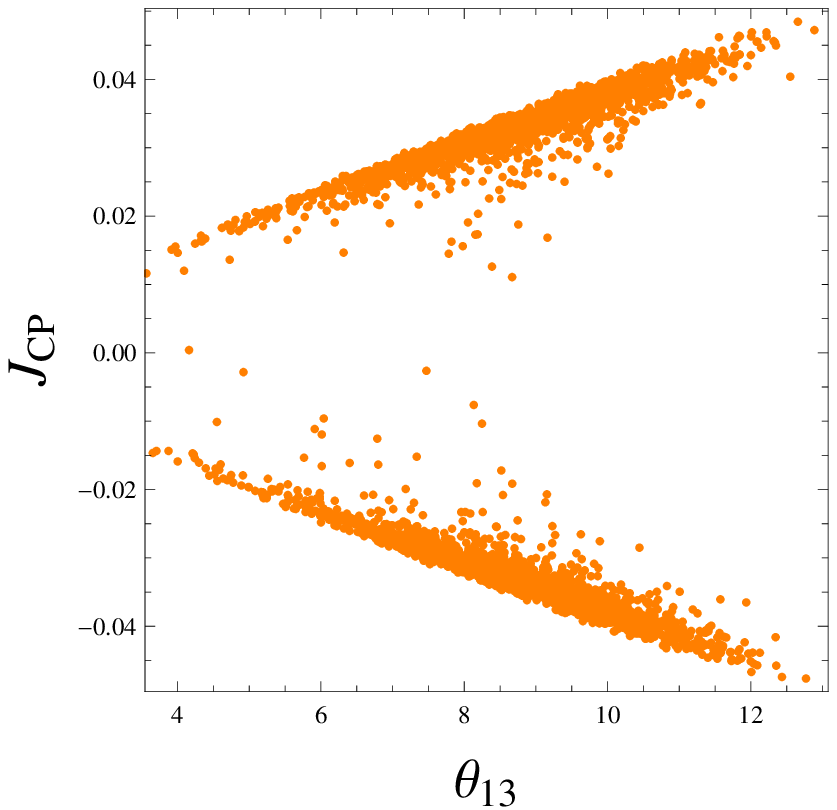, width=4.0cm, height=4.0cm} \epsfig{file=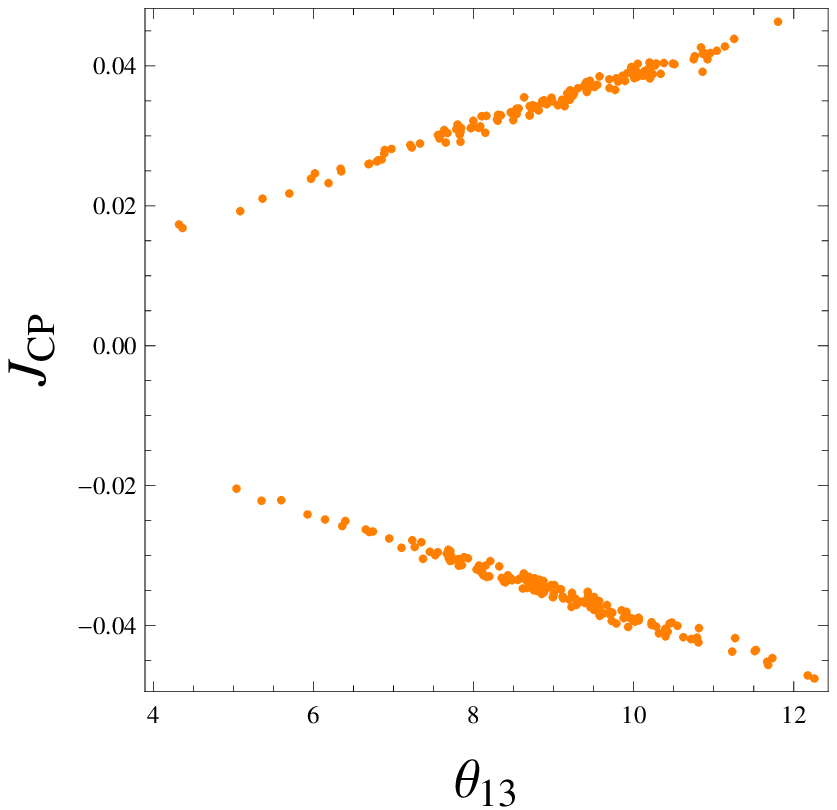, width=4.0cm, height=4.0cm} \epsfig{file=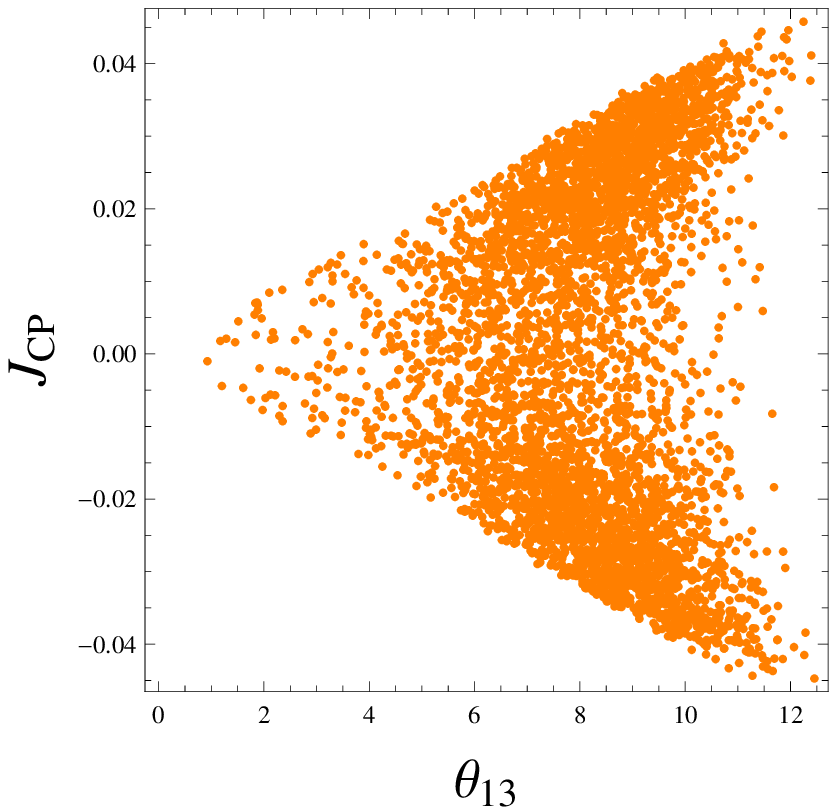, width=4.0cm, height=4.0cm}}
{\epsfig{file=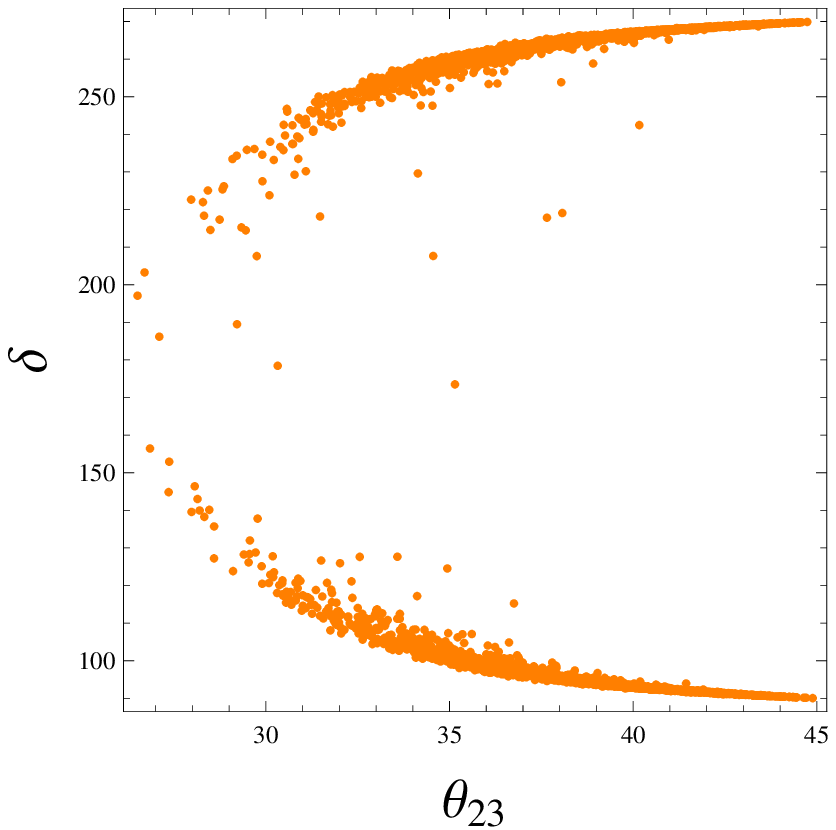, width=4.0cm, height=4.0cm} \epsfig{file=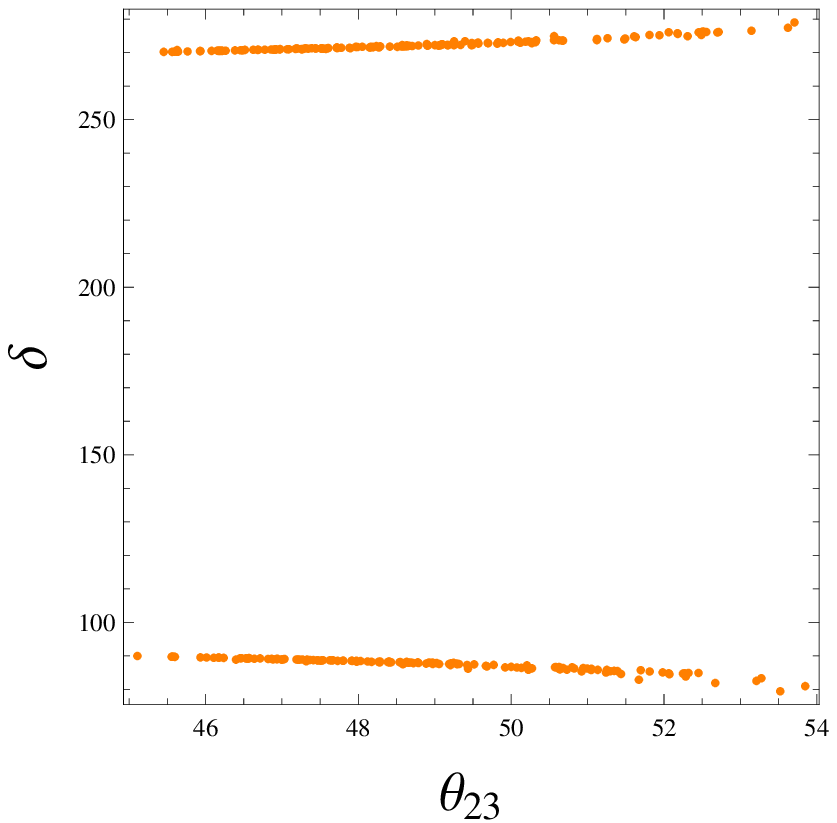, width=4.0cm, height=4.0cm} \epsfig{file=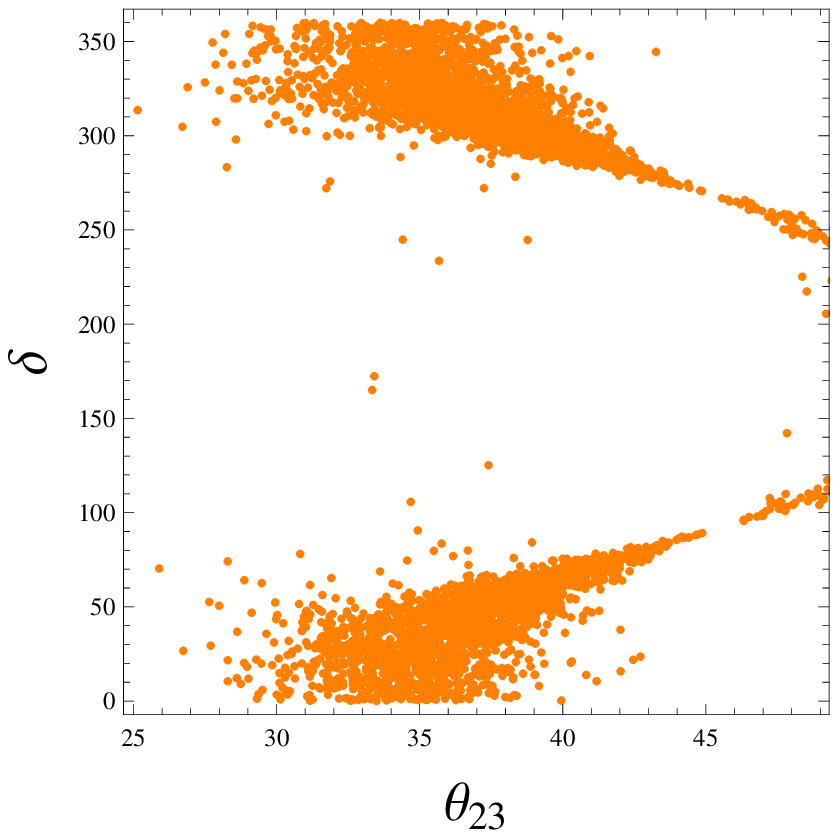, width=4.0cm, height=4.0cm}}
\end{center}
\caption{The correlation plots of $\theta_{13}$ (in degrees) and $\theta_{23}$ (in degrees) with $J_{CP}$ and $\delta$ (in degrees), respectively, for normal hierarchy (NH) of neutrino masses. The first column is for type I neutrino mass model, second for type II and third for type III neutrino mass model.}
\end{figure}

\begin{figure}
\begin{center}
{\epsfig{file=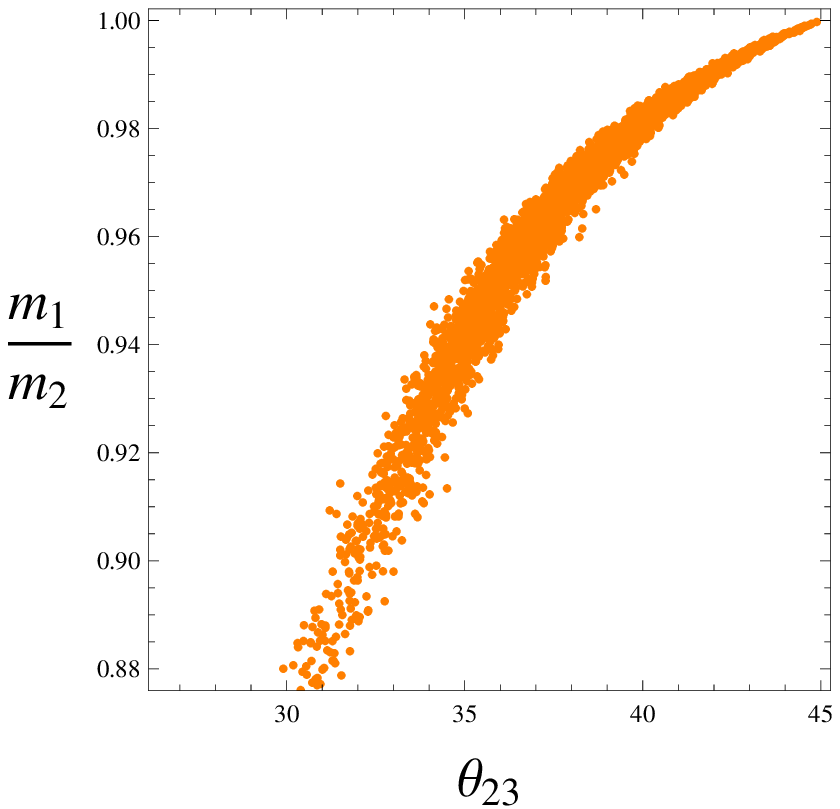, width=4.0cm, height=4.0cm} \epsfig{file=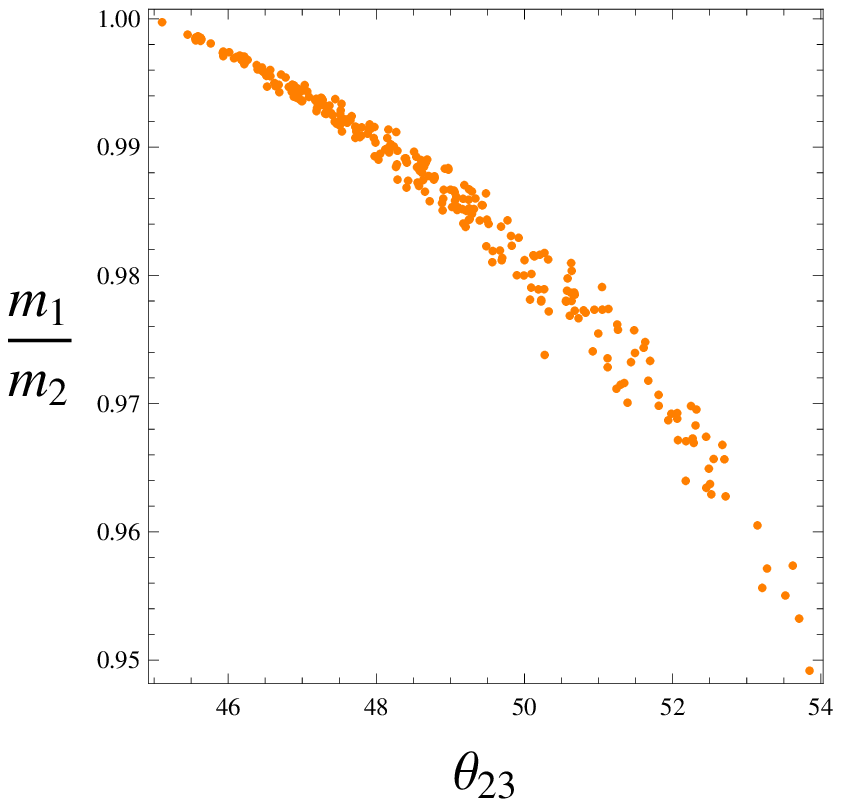, width=4.0cm, height=4.0cm} \epsfig{file=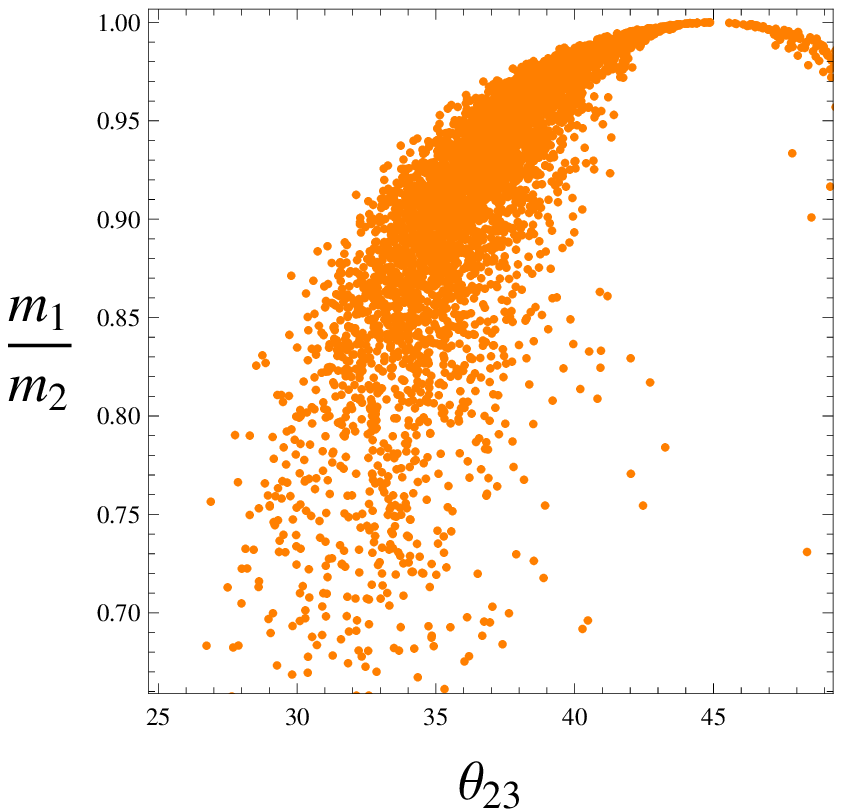, width=4.0cm, height=4.0cm}}

{\epsfig{file=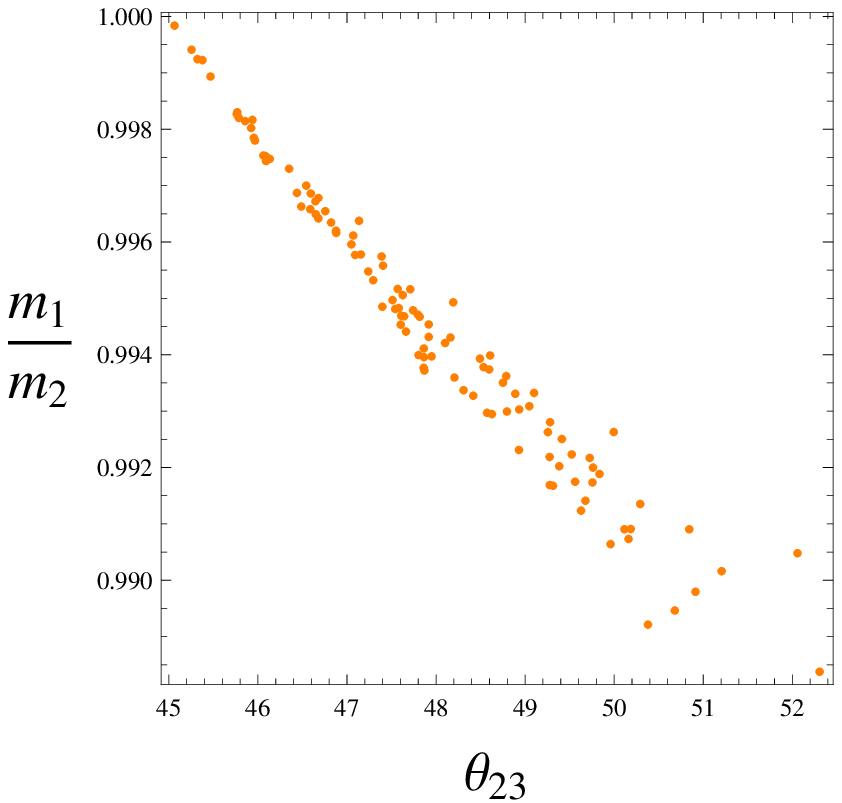, width=4.0cm, height=4.0cm} \epsfig{file=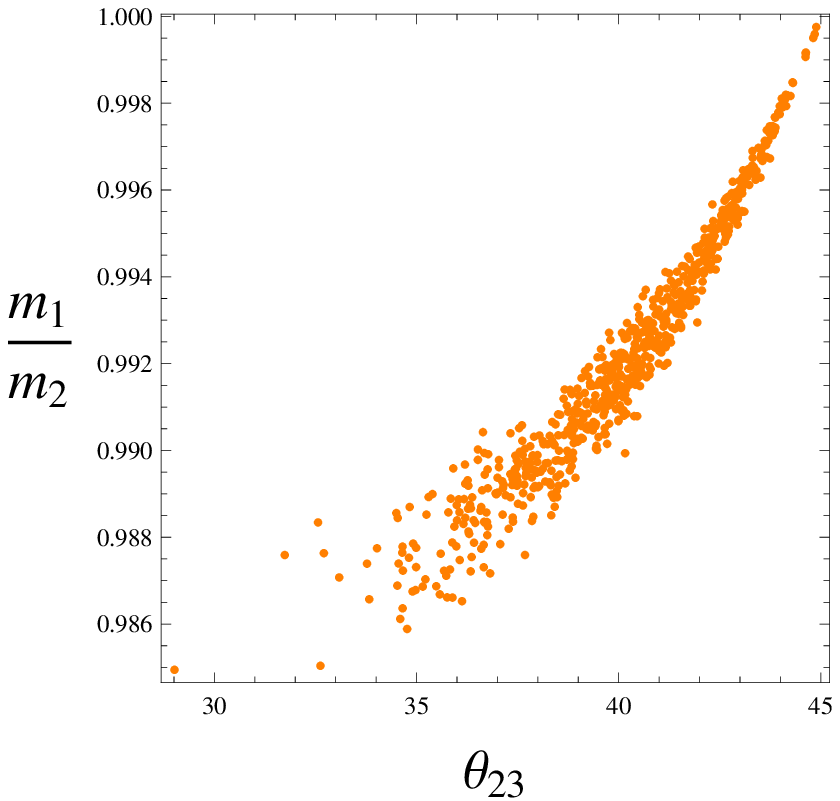, width=4.0cm, height=4.0cm} \epsfig{file=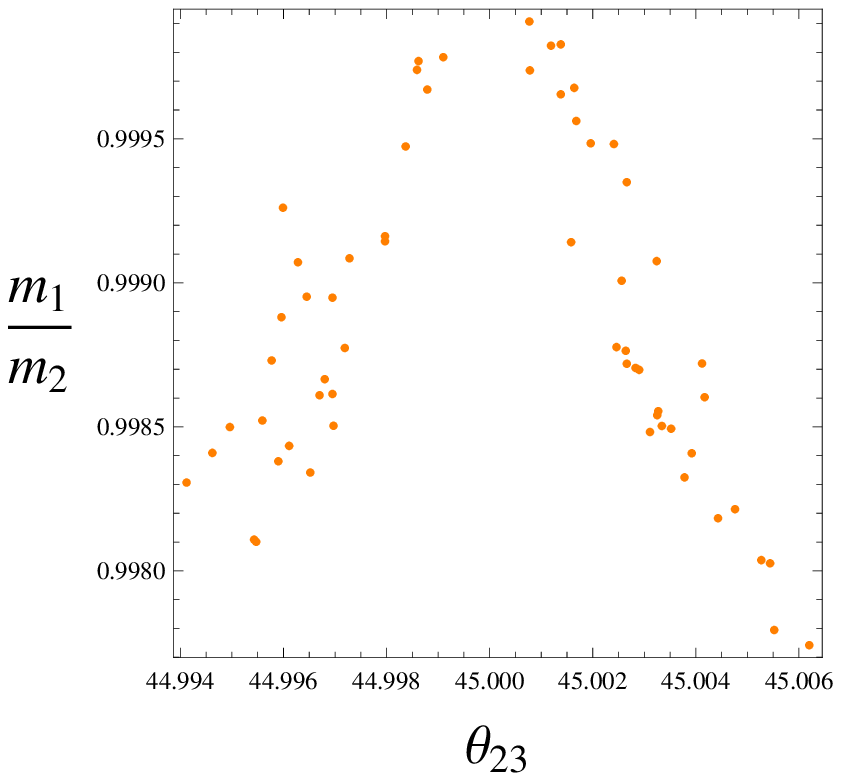, width=4.0cm, height=4.0cm}}
\end{center}
\caption{The correlation plots of $\theta_{23}$ (in degrees) with $\frac{m_1}{m_2}$ for normal (first row) as well as inverted (second row) hierarchy of neutrino masses. The first column is for type I neutrino mass model, second for type II and third for type III neutrino mass model.}
\end{figure}

\begin{figure}
\begin{center}
{\epsfig{file=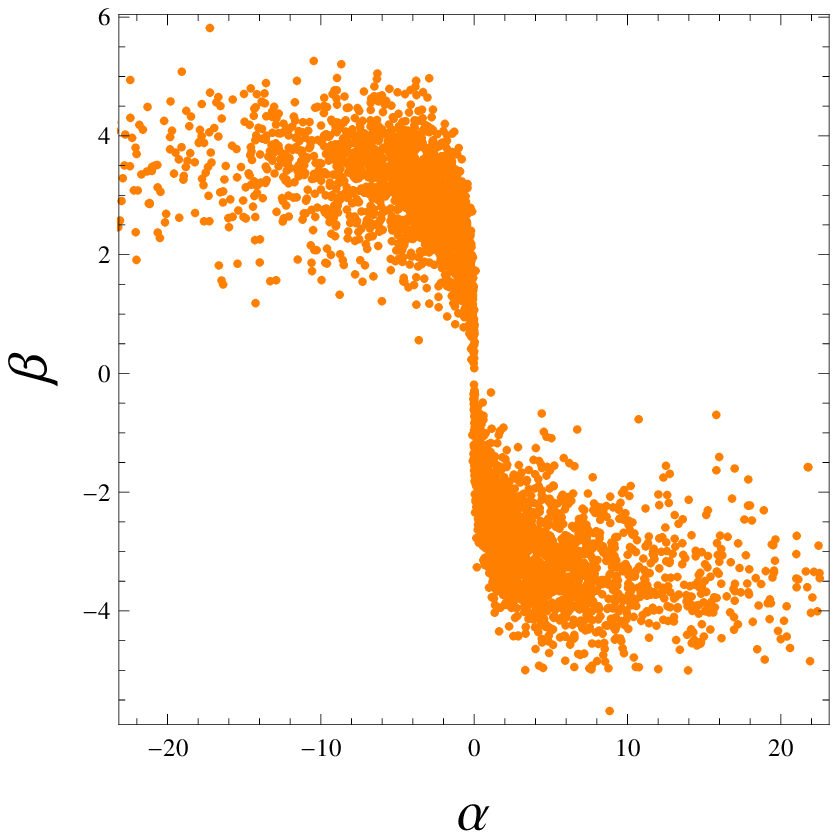, width=4.0cm, height=4.0cm} \epsfig{file=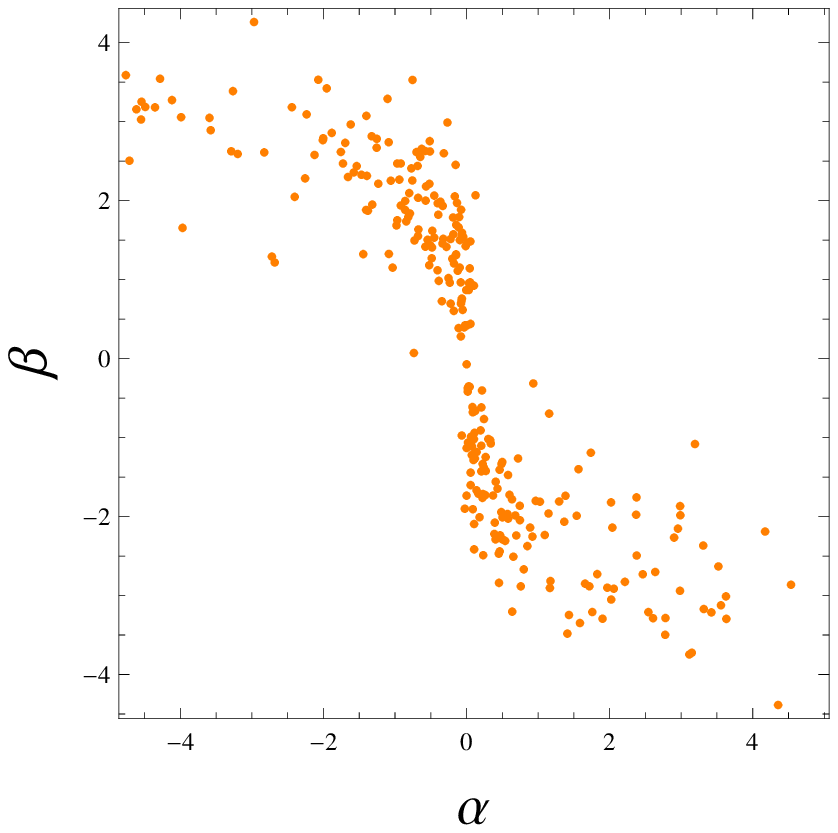, width=4.0cm, height=4.0cm} \epsfig{file=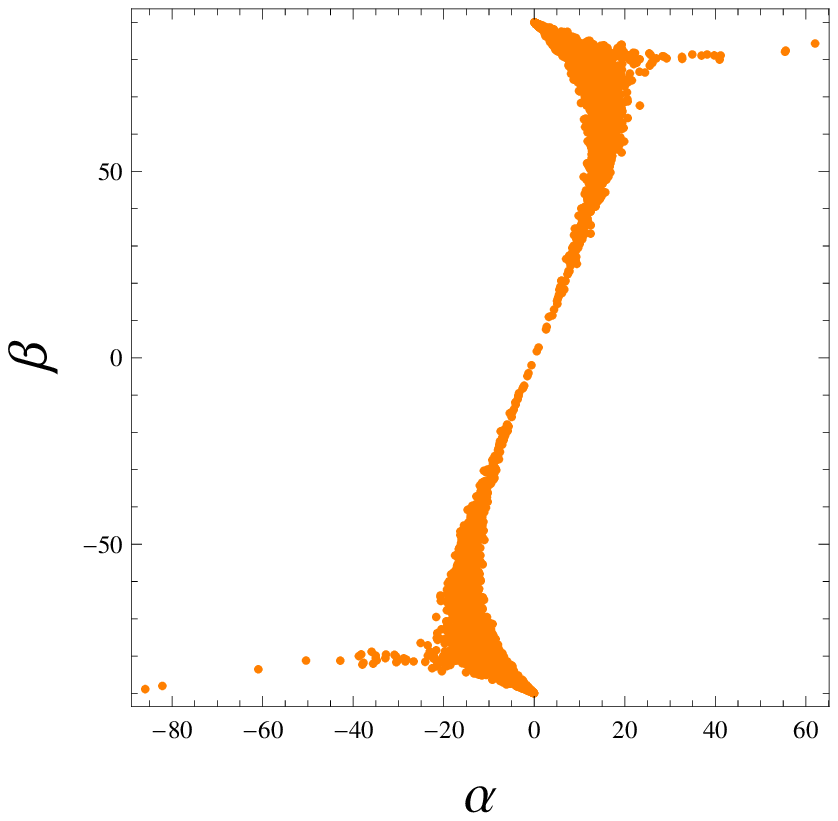, width=4.0cm, height=4.0cm}}

{\epsfig{file=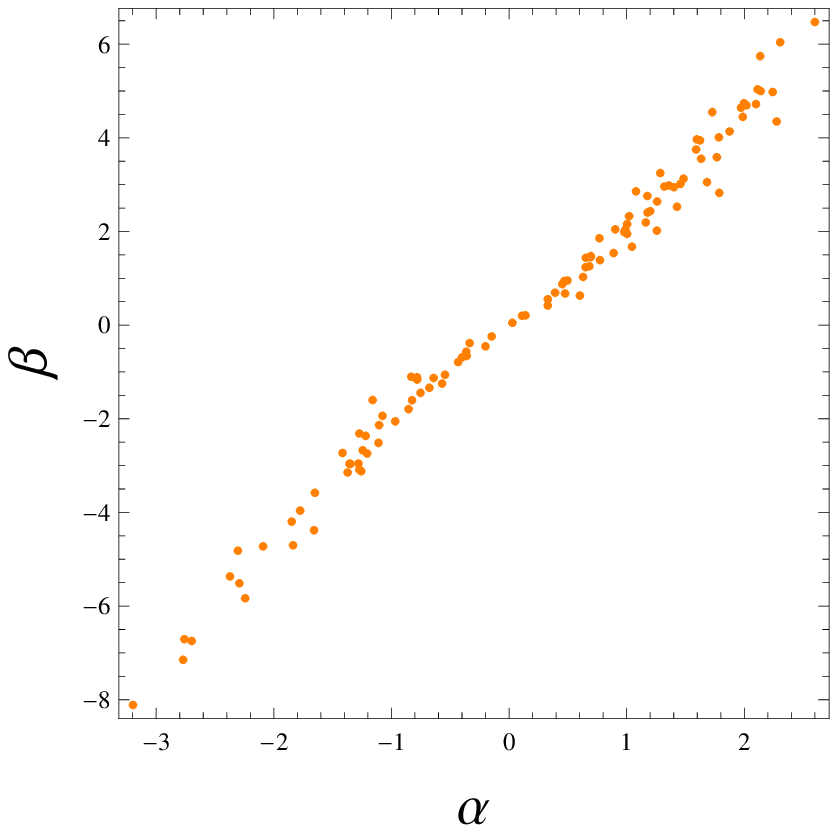, width=4.0cm, height=4.0cm} \epsfig{file=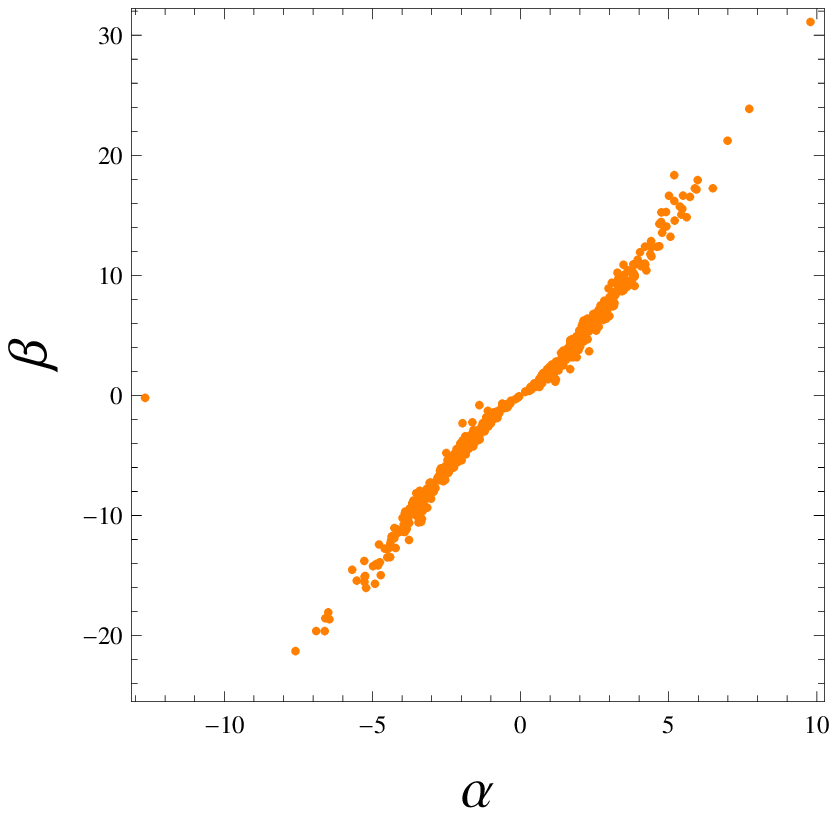, width=4.0cm, height=4.0cm} \epsfig{file=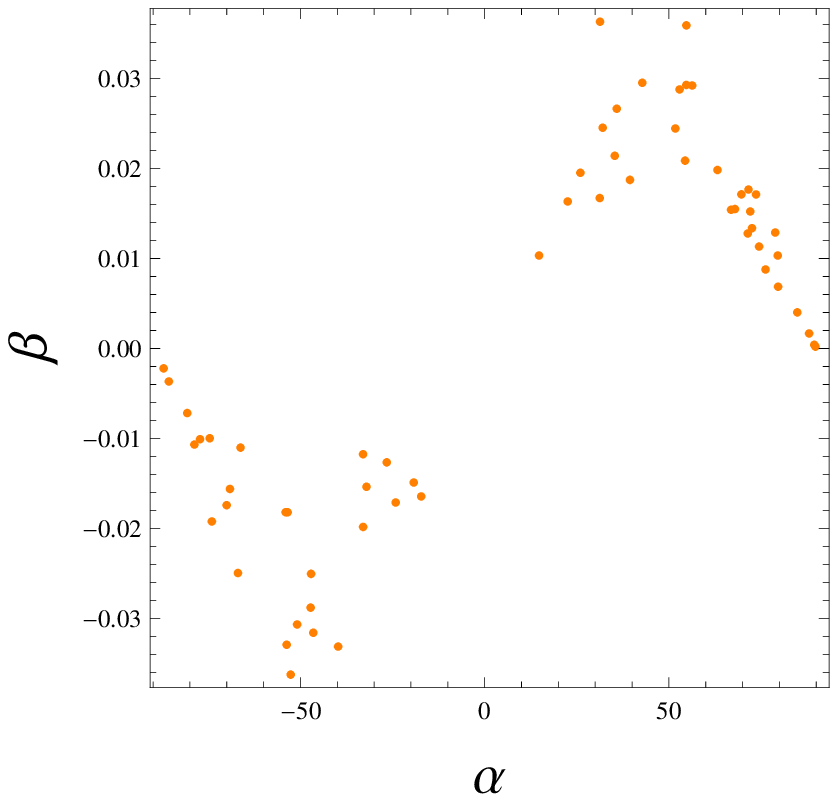, width=4.0cm, height=4.0cm}}
\end{center}
\caption{The correlation plots of the Majorana phases $\alpha$ (in degrees), $\beta$ (in degrees) for normal (first row) and inverted (second row) hierarchy. The first column is for type I neutrino mass model, second for type II and third for type III neutrino mass model.}
\end{figure}

\begin{figure}
\begin{center}
 {\epsfig{file=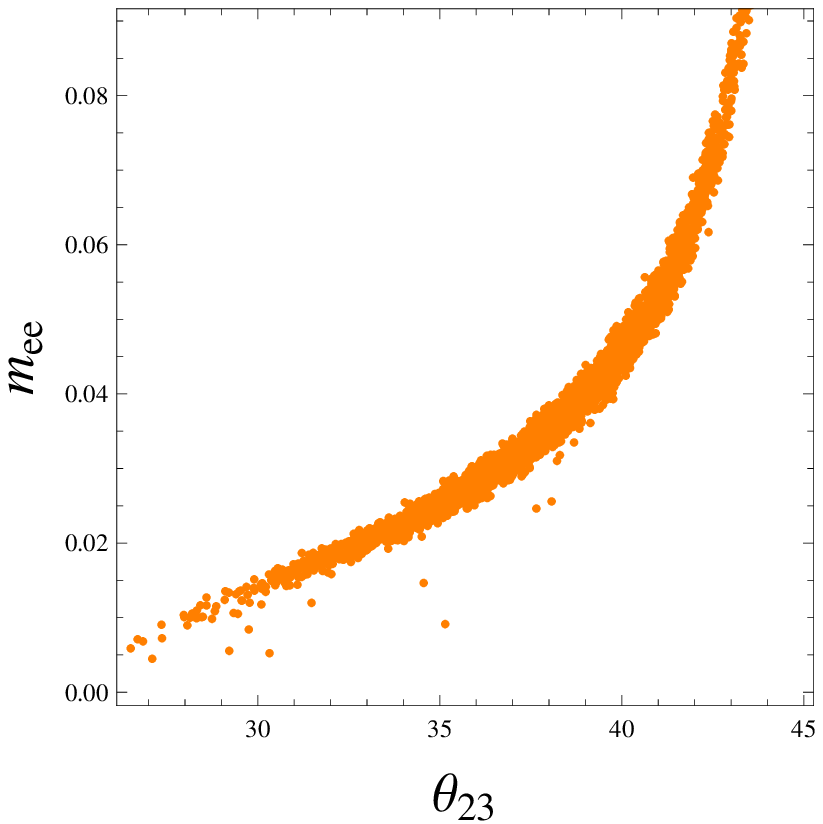, width=5.0cm, height=5.0cm} \epsfig{file=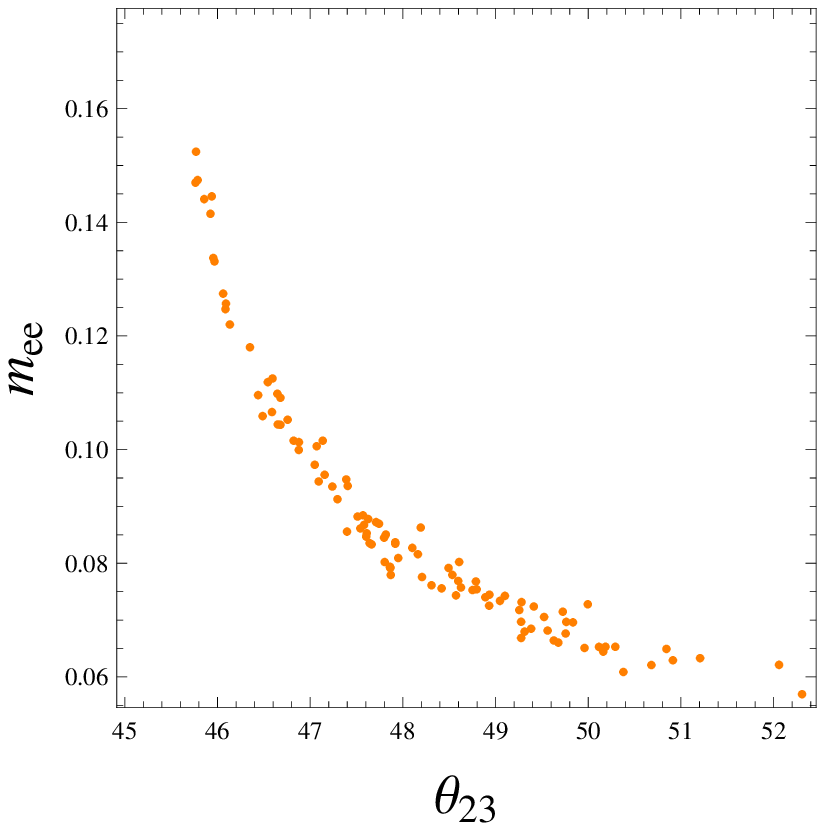, width=5.0cm, height=5.0cm}} 
{\epsfig{file=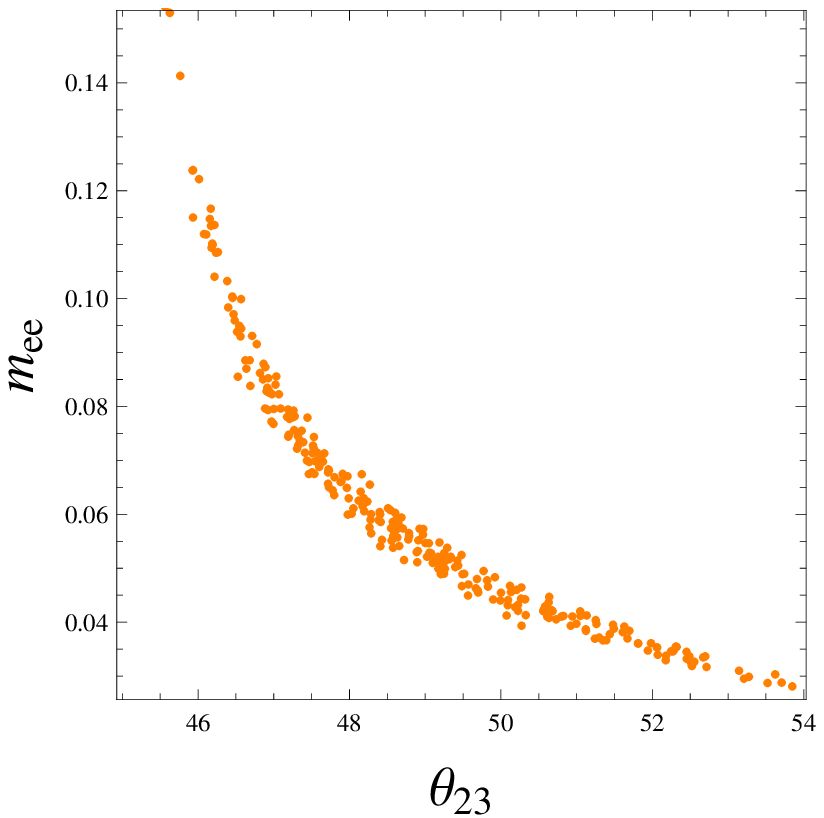, width=5.0cm, height=5.0cm} \epsfig{file=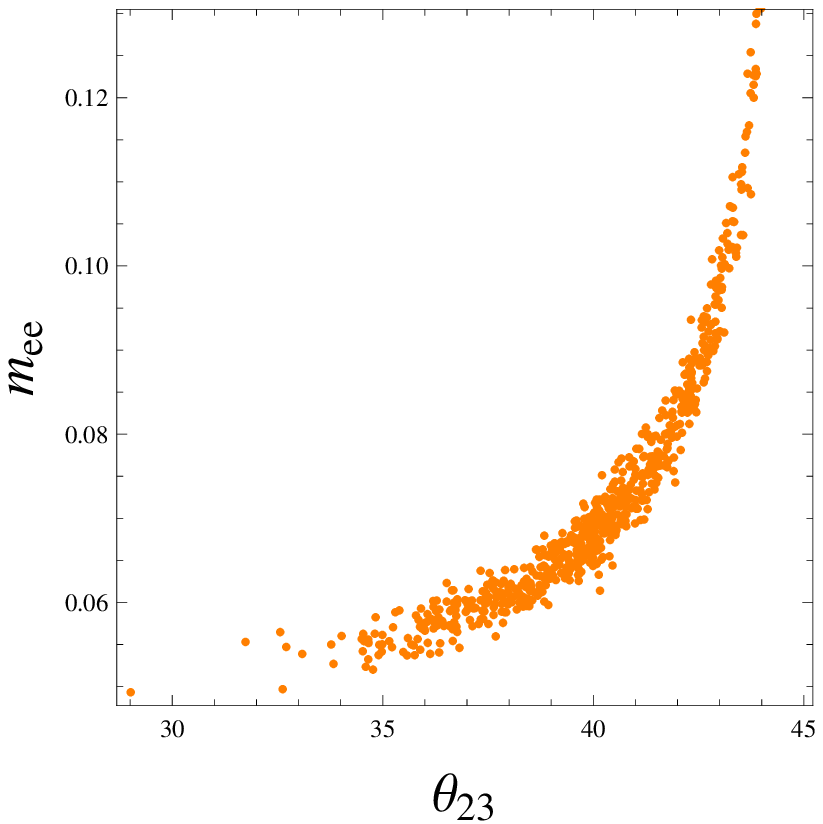, width=5.0cm, height=5.0cm}} 
{\epsfig{file=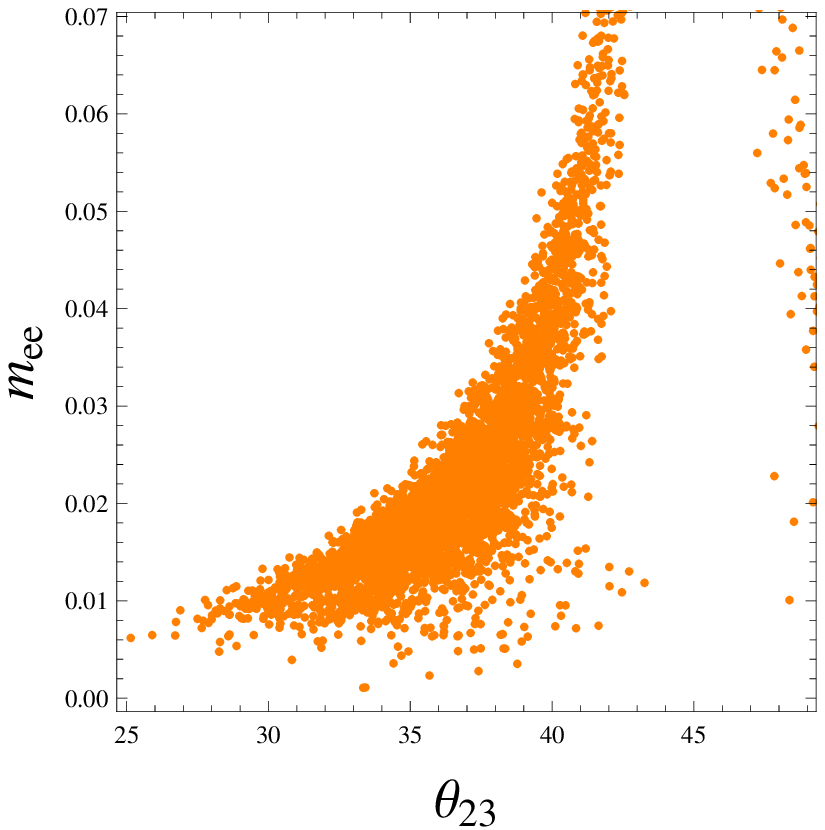, width=5.0cm, height=5.0cm}\epsfig{file=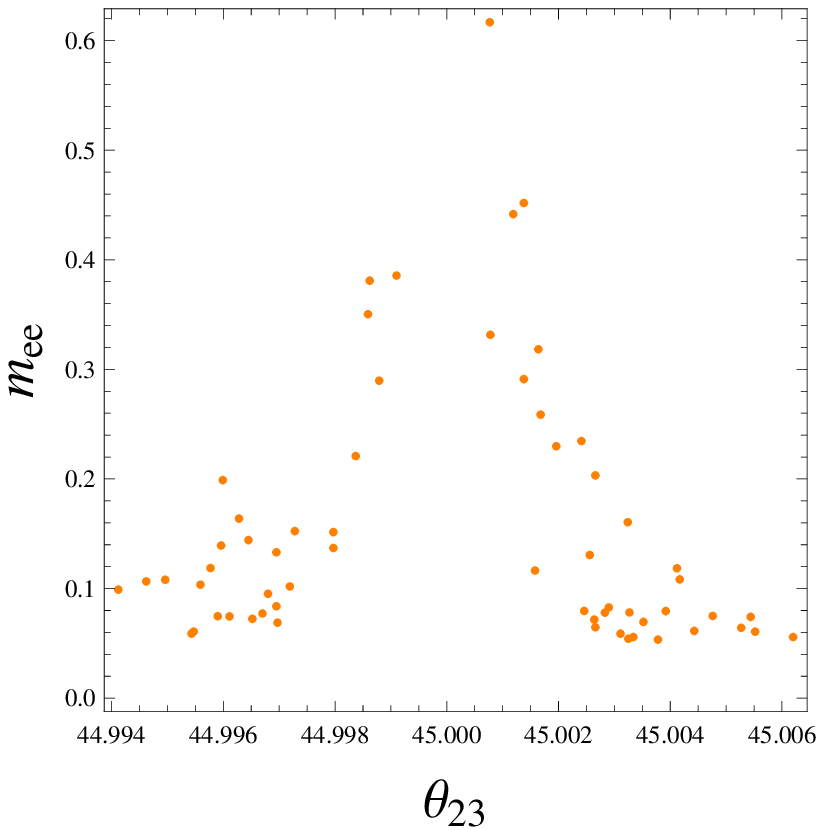, width=5.0cm, height=5.0cm}}
\end{center}
\caption{The correlation plots of effective Majorana mass $m_{ee}$ (in eV) with $\theta_{23}$ (in degrees) for normal (first column) and inverted (second column) hierarchy. The first row depicts the correlation for type I neutrino mass model, second for type II and third for type III neutrino mass model.}
\end{figure}

\end{document}